\documentclass[a4paper,11pt]{article}
\usepackage{pos}
\usepackage{sidecap}
\sidecaptionvpos{figure}{t}

\title{\emph{Trinity}'s Sensitivity to Isotropic and Point-Source Neutrinos}
 \ShortTitle{\emph{Trinity}'s Sensitivity to Isotropic and Point-Source Neutrinos}

\author*[a]{Andrew Wang}
\author[a]{Chaoxian Lin}
\author[a]{Nepomuk Otte}
\author[b]{Michele Doro}
\author[a]{Eliza Gazda}
\author[a]{Ignacio Taboada}
\author[c]{Anthony M. Brown}
\author[a]{Mahdi Bagheri}

\affiliation[a]{Georgia Institute of Technology, School of Physics \& Center for Relativistic Astrophysics,\\
837 State Street NW, Atlanta, Georgia 30332-0430, USA}

\affiliation[b]{Università di Padova (UniPD), Dipartimento di Fisica e Astronomia (DFA) G. Galilei \\
I-35131 Padova, Italy
}

\affiliation[c]{Centre for Advanced Instrumentation, Durham University,
  South Road, Durham, DH1 3LE, UK}

\emailAdd{a.w@gatech.edu}

\abstract{
The neutrino band above 10\,PeV remains one of the last multi-messenger windows to be opened, a challenge that several groups tackle. One of the proposed instruments is \emph{Trinity}, a system of air-shower imaging telescopes to detect Earth-skimming neutrinos with energies from $10^6$\,GeV to $10^{10}$\,GeV. We present updated sensitivity calculations demonstrating \emph{Trinity}'s capability of not only detecting the IceCube measured diffuse astrophysical neutrino flux but doing so in an energy band that overlaps with IceCube's. \emph{Trinity} will distinguish between different cutoff scenarios of the astrophysical neutrino flux, which will help identify their sources. We also discuss \emph{Trinity}'s sensitivity to transient sources on timescales from hours to years.}

\FullConference{37$^{\rm{th}}$ International Cosmic Ray Conference (ICRC 2021)\\
		July 12th -- 23rd, 2021\\
		Online -- Berlin, Germany}

\begin{document}

\maketitle

\section{Introduction}

Neutrino astrophysics is a rapidly developing field. A significant step forward has been the detection of astrophysical neutrinos by the IceCube team \cite{ICDetection2013}. The opening of the high-energy neutrino band has renewed the interest in the detection of neutrinos at even higher energies with ultra-high energy (UHE, $>10$\,PeV) neutrinos. UHE neutrinos shed light on the unknown origin of IceCube's astrophysical neutrinos and help answer other pressing questions in neutrino astrophysics, cosmic-ray physics, and fundamental neutrino physics. 

While UHE neutrinos must exist, their detection poses a formidable intellectual challenge tackled with several complementary approaches. One of the proposed techniques and pursued with the proposed \emph{Trinity} UHE-neutrino observatory is the imaging of air-showers initiated by Earth-skimming UHE neutrinos. For \emph{Trinity}, we present updated diffuse neutrino flux and point source sensitivities. After a description of \emph{Trinity}, we discuss how we calculate \emph{Trinity}'s sensitivity before presenting our results and discussing them in the context of TXS 0506+056 and NGC 1068.

\section{Conceptual Design of the Trinity Observatory}

\begin{figure}
    \centering
    \includegraphics[width=0.9\textwidth]{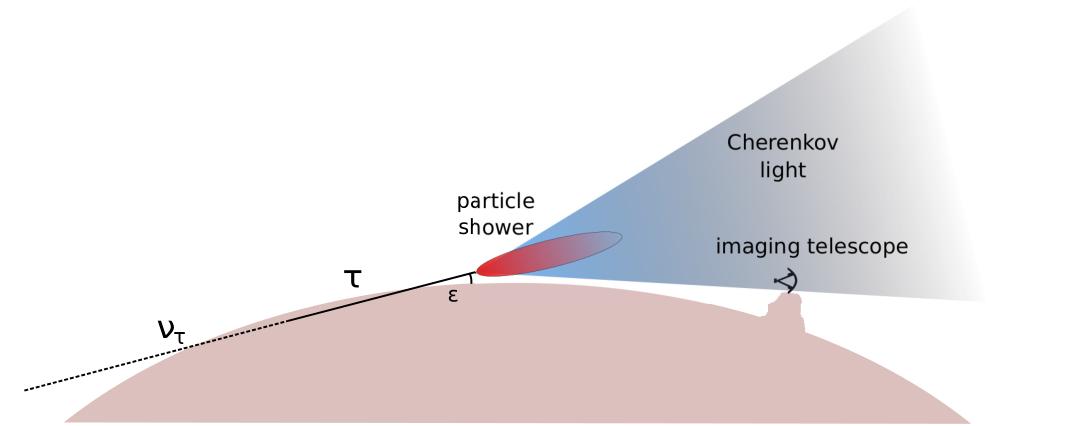}
    \caption{Simplified diagram of the Earth-skimming technique employed by \emph{Trinity} to detect UHE tau neutrinos. A UHE tau neutrino enters the Earth and interacts, prompting the production of a tau that emerges from the Earth and decays into a particle shower. The Cherenkov light produced by the shower particles can then be detected by \emph{Trinity}'s imaging system situated at an altitude of 2\,km.}
    \label{fig:skim}
\end{figure}

UHE neutrinos that enter the Earth interact within tens to hundreds of kilometers. Therefore, the tau produced in the interaction of a tau neutrino skimming the Earth, as shown in Figure \ref{fig:skim}, has a high probability of emerging from the ground. The tau then decays, and an extensive particle shower develops in the lower atmosphere. 

The majority of the shower particles emit Cherenkov light, which is intense enough such that an air-shower imaging system collects enough light to image the air shower at a distance of up to 200\,km \cite{Otte2019}. For air showers developing within tens of kilometers from the telescope, the air shower can also be imaged with the fluorescent light emitted from collision excited atmospheric nitrogen molecules. The \emph{Trinity} UHE-neutrino observatory is a system of air-shower imaging telescopes optimized to detect these upward-going air showers. 

In its proposed configuration, which is also the configuration assumed here, the \emph{Trinity} telescopes are at three sites at an altitude of about 2\,km. At each location, all telescopes point at the horizon and provide a combined $360^\circ$ azimuthal coverage. The telescopes' $5^\circ$ field-of-view in elevation ensures that air-shower images are not truncated. We assume a $0.3^\circ$ angular resolution for the optics and the camera and an effective light-collection area in any direction of 10\,m$^2$ \cite{ICRCAnthony,ICRCAnthony2021}. For the efficiency of the photon detectors, we use the characteristics of the blue-sensitive Hamamatsu S14520-6050CN silicon photomultiplier (SiPMs). The sensitivity of \emph{Trinity} scales linearly with the volume of atmosphere monitored. Because the fields of view of Trinity's telescopes do not overlap, the sensitivity thus scales proportionally with the number of telescopes.

\section{Sensitivity Calculations}

We follow the same approach for the sensitivity calculation as described in \cite{Otte2019}. However, we replaced the analytic calculation of the neutrino interaction and tau propagation in the Earth with lookup tables generated with NuTauSim \cite{Alvarez-Muniz2018a}. NuTauSim better describes the tau energy loss and includes tau-neutrino regeneration. The updated tau propagation model results in a significantly better sensitivity below $10^8$\,GeV than our previous calculations \cite{Otte2019}. We could trace back the discrepancy to an error in the parameterization of the energy loss for low-energy taus in our earlier calculations. 
 
The sensitivity is defined as the flux $\Phi$ resulting in one detected neutrino within the effective observation time $T$ for an assumed power-law spectrum with a spectral index $\gamma$:
\begin{equation}
    \Phi(E_\nu) = \dfrac{3 E_{\nu}^{-\gamma}}{T \times \int_{E_{\nu ,\text{min}}}^{E_{\nu ,\text{max}}}A\left(E_{\nu'}\right) E_{\nu'}^{-\gamma}dE_{\nu'}}
\end{equation}
where $E_\nu$ is the central neutrino energy, $E_{\nu ,\text{min}}$ and $E_{\nu ,\text{max}}$ are the minimum and maximum observable neutrino energies in the energy bin, and $A$ is the effective area or acceptance of \emph{Trinity}. The factor three converts the tau-neutrino-only flux into a 1:1:1 flavor mixed flux commonly assumed for astrophysical fluxes arriving at Earth.

\subsection{Diffuse Flux Sensitivity}
For the diffuse-flux sensitivity, we assume a ten-year-long observation with a duty cycle of 20\%. A slightly lower duty cycle of 18\% is regularly demonstrated with very-high-energy gamma-ray telescopes, which have more stringent requirements to observe under clear skies.

We calculate the acceptance for a diffuse neutrino flux as
\begin{equation}
    A(E_\nu,\epsilon) = \int_{A'} \int_\Omega \int_{E_\tau} P(E_\tau | E_\nu , \epsilon) \cdot \sin{(\epsilon)} dE_\tau dA' d\Omega
\end{equation}
where $\epsilon$ is the angle of incidence of the neutrino relative to the ground. The outer two integrals are over the surface area and all possible arrival directions of neutrinos. The inner integral is over tau energy. $P$ is the combined probability that a) a neutrino with a given energy and direction interacts in the Earth, and a tau emerges from the ground, b) the tau decays, and an air shower develops fully within the field of view of \emph{Trinity}, c) the telescope receives enough light to image the air shower and later reconstruct the event, and d) the length of the air shower is more than $0.3^\circ$. 
 
\begin{figure}[ht]
\includegraphics*[trim = 0 0 40 60, clip, angle=0, width=0.495\textwidth]{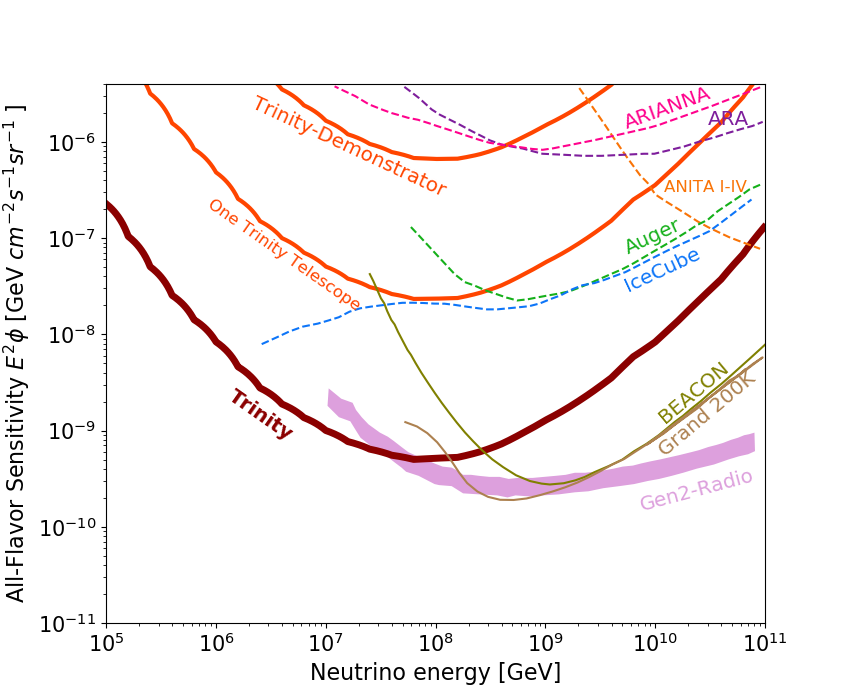}
\hfill
\includegraphics*[trim = 0 0 40 60, clip, angle=0, width=0.495\textwidth]{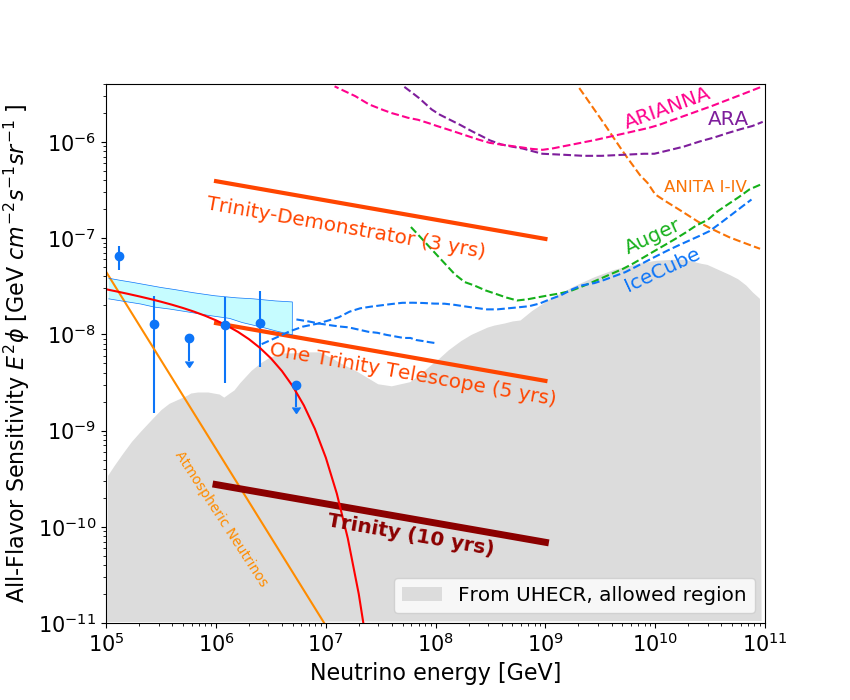}
\caption{\emph{Trinity}'s differential sensitivity (left) and integral sensitivity (right). \emph{Trinity} Demonstrator is a single 1\,m$^2$ telescope scheduled to operate for three years. The middle orange curve is for one \emph{Trinity} telescope operating for five years. The dark red curve shows the ten-year sensitivity of the complete \emph{Trinity} Observatory. The dashed curves represent published upper limits on the diffuse UHE neutrino flux. The integral sensitivity curves assume a power law with a spectral index of $-2$. Gray shaded is the area of predicted fluxes. The blue data points and the blue shaded bow tie are IceCube's measurements and spectral fit, respectively, of the astrophysical neutrino flux. Figures adapted from \cite{ICG22021}.}
\label{fig:sensitivity}
\end{figure}
 
Figure~\ref{fig:sensitivity} left shows the differential sensitivity for three different development stages of \emph{Trinity}. For these calculations, we integrated the acceptance over one order of magnitude in energy. The upper-most orange curve denotes the three-year sensitivity of a 1\,m$^2$ \emph{Trinity}-Demonstrator telescope with a $5^\circ \times 5^\circ$ field of view. The orange curve in the middle shows the five-year sensitivity of one \emph{Trinity} telescope with a $60^\circ$ azimuthal field of view. The lowest, dark red curve shows the ten-year sensitivity of the complete \emph{Trinity} observatory with three sites.  

Figure~\ref{fig:sensitivity} right shows the integral sensitivity from $10^6$\,GeV to $10^{9}$\,GeV. The energy range covers the interval over which the central 90\% of neutrinos would be detected for a power-law spectrum with index $-2.2$. The blue data points are IceCube's measurements of the astrophysical neutrino flux. The spectral index assumed in our integral sensitivity calculations matches the index of the astrophysical neutrino spectrum. The gray shaded area shows the phase space of predicted neutrino fluxes.

\begin{SCfigure}[1.0][!b]
\includegraphics*[angle=0,width=0.55\textwidth]{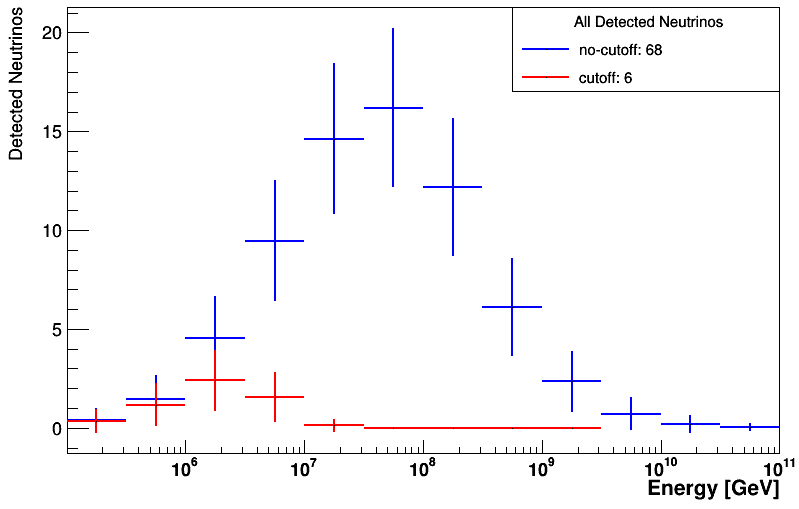}
\caption{Simulated distributions of astrophysical neutrinos detected with \emph{Trinity} over a ten-year long observational for two different scenarios. The blue distribution assumes a power-law spectrum for the astrophysical neutrinos without a cutoff. The red distribution assumes the same spectrum but applies an exponential cutoff constrained by the IceCube upper limit at $5 \times 10^6$ GeV (solid red curve, Figure~\ref{fig:sensitivity}, right).}
\label{fig:cutoff}
\end{SCfigure}

A unique feature of \emph{Trinity} is its low energy threshold of $10^6$\,GeV, which enables \emph{Trinity} to detect the astrophysical neutrino flux at energies that overlap with IceCube's measurements. More importantly, \emph{Trinity} can extend IceCube's measurement to higher energies and essential astrophysical knowledge is gained about the origin of the flux. \emph{Trinity}'s ability to discriminate between different flux scenarios illustrates Figure~\ref{fig:cutoff} on the example of two extreme cases. In case the spectrum extends without a break, \emph{Trinity} will detect 70 neutrinos, corresponding to seven neutrino detections per year. Even if the spectrum breaks off, as shown in Figure \ref{fig:sensitivity} right, \emph{Trinity} will still detect six neutrinos in ten years. Any other number of neutrinos detected will be a sensitive measure of a spectral cutoff that falls in between these two extremes. Thus, even without reconstructing the energy of \emph{Trinity} detected events, we gain new information about the spectral shape simply by joining IceCube and \emph{Trinity} results. The energy overlap with IceCube a) ensures that astrophysical neutrinos will be a guaranteed signal for \emph{Trinity} and b) opens up the unique possibility to cross calibrate the two instruments.

\emph{Trinity}'s energy overlap with IceCube and its exclusive sensitivity to tau neutrinos will, furthermore, enable a joint analysis of both experiments that helps constrain the flavor flux ratio of astrophysical neutrinos, thus probing the environment around the neutrino sources such as the magnetic field \cite{Kashti2005}. Beyond standard astrophysical processes and neutrino oscillations, joint studies utilizing flavor information will probe physics beyond the standard model, such as neutrino decay \cite{Bustamante2017} or non-standard neutrino oscillations \cite{Bustamante2015}.

\subsection{Point-Source Sensitivity}

\begin{figure}[th]
\includegraphics*[width=0.495\textwidth]{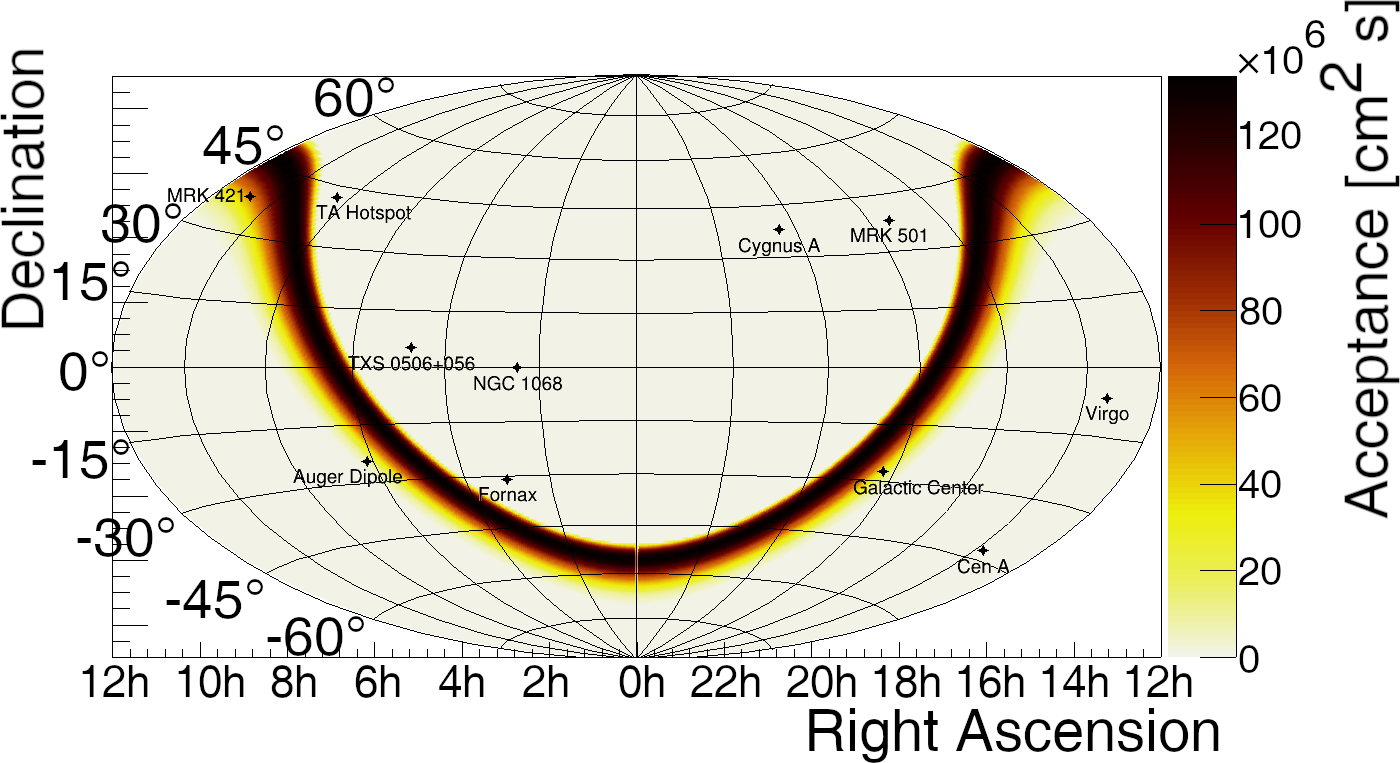}
\hfill
\includegraphics*[width=0.495\textwidth]{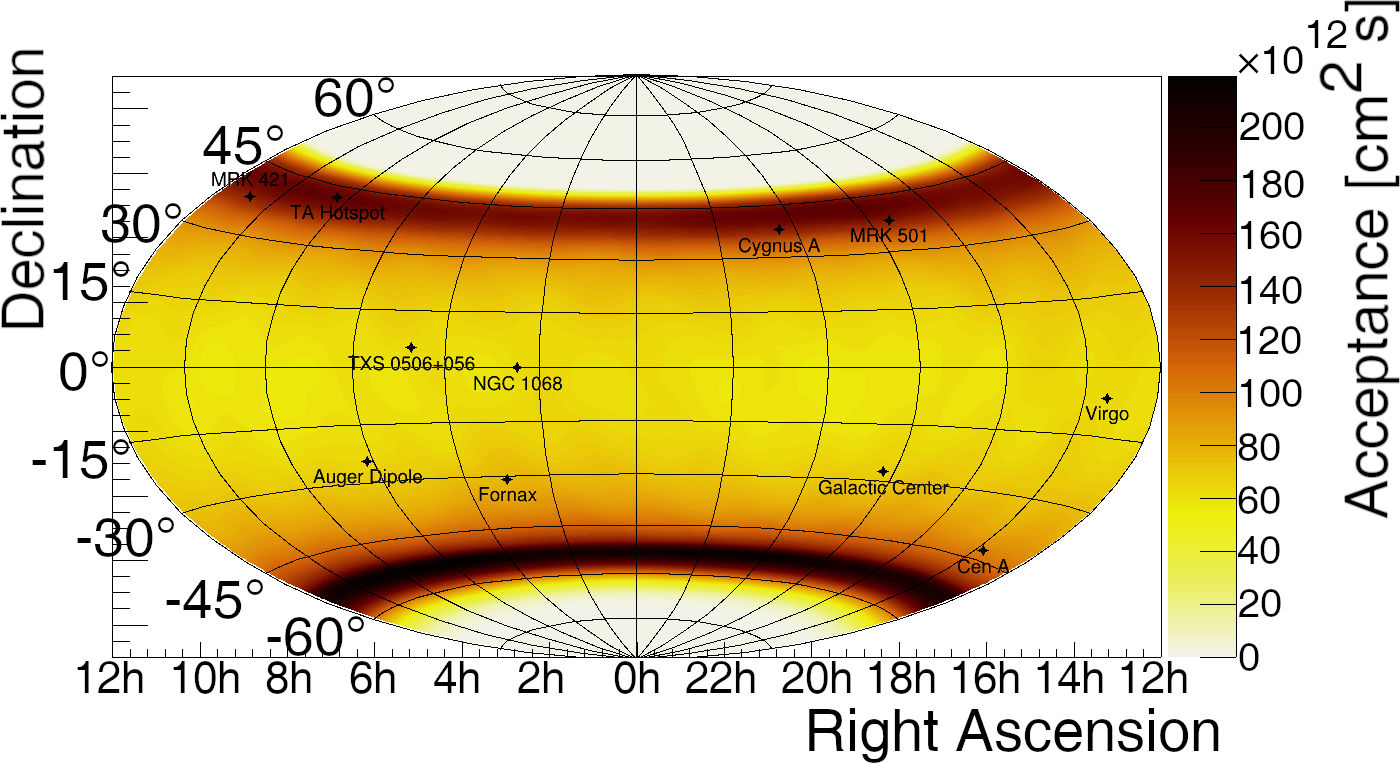}
\caption{Left: Instantaneous acceptance of one \emph{Trinity} site projected onto the celestial sphere. Right: 1-year integrated exposure of one \emph{Trinity} site. Overlayed are relevant astrophysical sources within the field of view of \emph{Trinity}.}
\label{fig:accskymaps}
\end{figure}

We calculated the instantaneous acceptance projected back at the sky for the point-source sensitivity using the same code we used to calculate the diffuse flux sensitivity. Figure \ref{fig:accskymaps}, left, shows the instantaneous acceptance from $10^6$\,GeV to $10^{10}$\,GeV to a neutrino spectrum with a $-2$ index power law and assuming only one of the three \emph{Trinity} sites. For the location, we pick Frisco Peak, UT with 360$^\circ$ azimuthal coverage. The combined acceptance of three sites would result in three independent acceptance circles projected at different locations in the sky.
The instantaneous acceptance band has a $10.1^\circ$  full width at half maximum (FWHM) and is maximal for elevations of $>10^\circ$ below the horizon in the observatory reference frame. 

To evaluate the sensitivity towards a specific source, we propagate the acceptance band across the sky as time progresses. We, furthermore, require that the sun is $14.5^\circ$ below the horizon and the moon phase is $<30$\% at the site. We time integrate the acceptance for each point in the sky whenever these conditions are fulfilled, and the acceptance is non-zero. 

The right panel in Figure \ref{fig:accskymaps} shows the exposure for each point in the sky after one year. Sources with declinations between $-70^\circ$ and $55^\circ$ declination are observable at the Frisco Peak, UT site ($38^\circ$ North latitude). The exposure is maximal at declinations of $-53^\circ$ and $40^\circ$. The annual observation time totals 1765 hours, resulting in a duty cycle of 20\%. The average time a source is observable during a single night is 4.8 hours and the most prolonged nightly observation period is 12 hours. 

\begin{figure}[th]
\includegraphics*[width=0.495\textwidth]{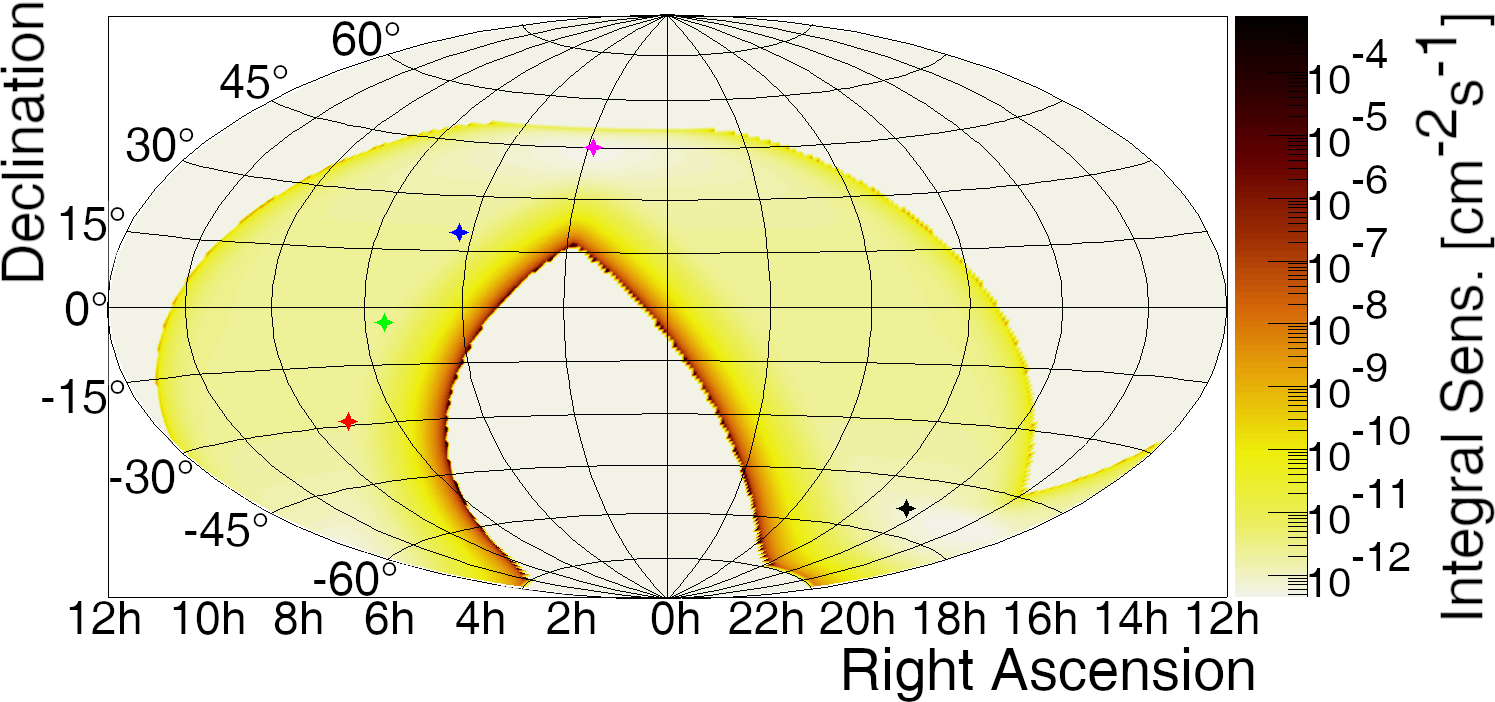}
\hfill
\includegraphics*[width=0.495\textwidth]{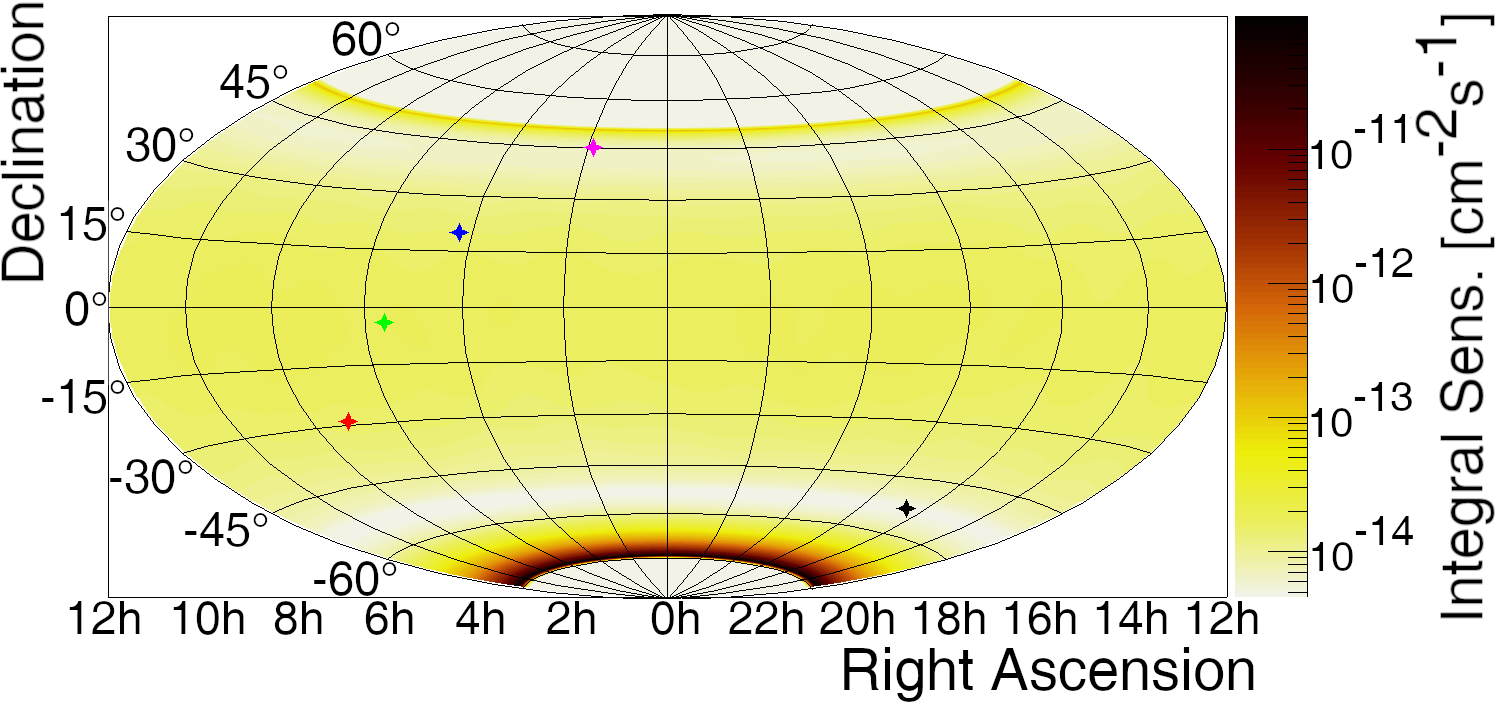}
\caption{Integral point-source sensitivity after five hours (left) and after one year (right). The stars depict source locations for which time dependent sensitivities are shown in Figure \ref{fig:ptsens}}
\label{fig:transitsens}
\end{figure}

To study \emph{Trinity}'s sensitivity to transient sources, we assumed two scenarios. In one case we assume a power-law spectrum with an index of $-2.2$ for the neutrino source and in the second case we assume a spectral index of $-3.2$. In both cases we require the detection of one neutrino as a definition of the minimal detectable flux. Figure~\ref{fig:transitsens}  left shows, as an example, the integral point-source sensitivity to a index $2.2$ source after five hours of observation and on the right after one year. As expected from the one-year acceptance shown in Figure~\ref{fig:accskymaps}, the sensitivity is highest at the highest and lowest accessible declination.

We evaluated \emph{Trinity}'s sensitivity to transient sources by simulating five sources distributed equally in declination between $-53^\circ$ to $45^\circ$ indicated by the stars in Figure \ref{fig:transitsens}. The right ascension of each location ensures that the source is immediately observed at the start of the simulation. For each source we assume fifteen flare durations between 1 hour and 1 year. Observational constraints like day/night cycles and the moon phases are taken into account.

\begin{figure}
\centering
\includegraphics[width=0.49\linewidth]{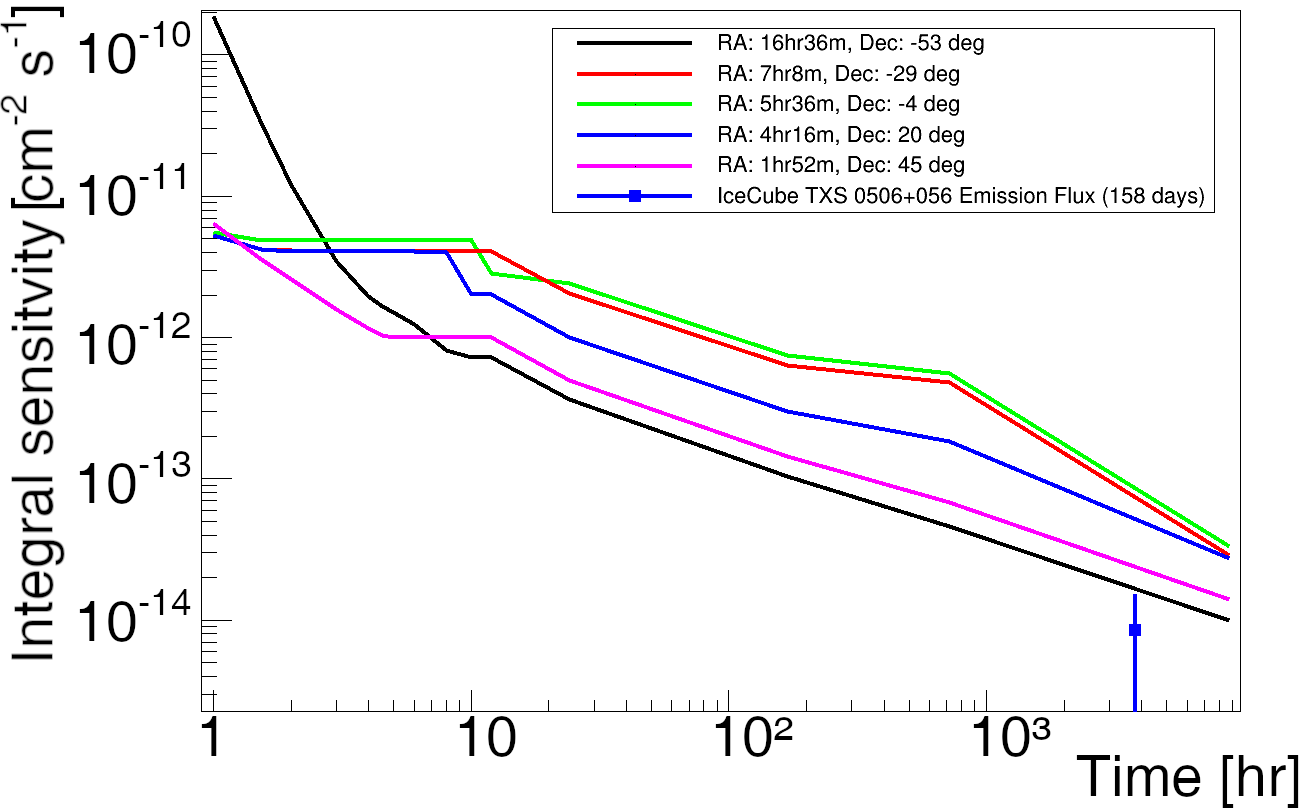}
\hfill
\includegraphics[width=0.49\linewidth]{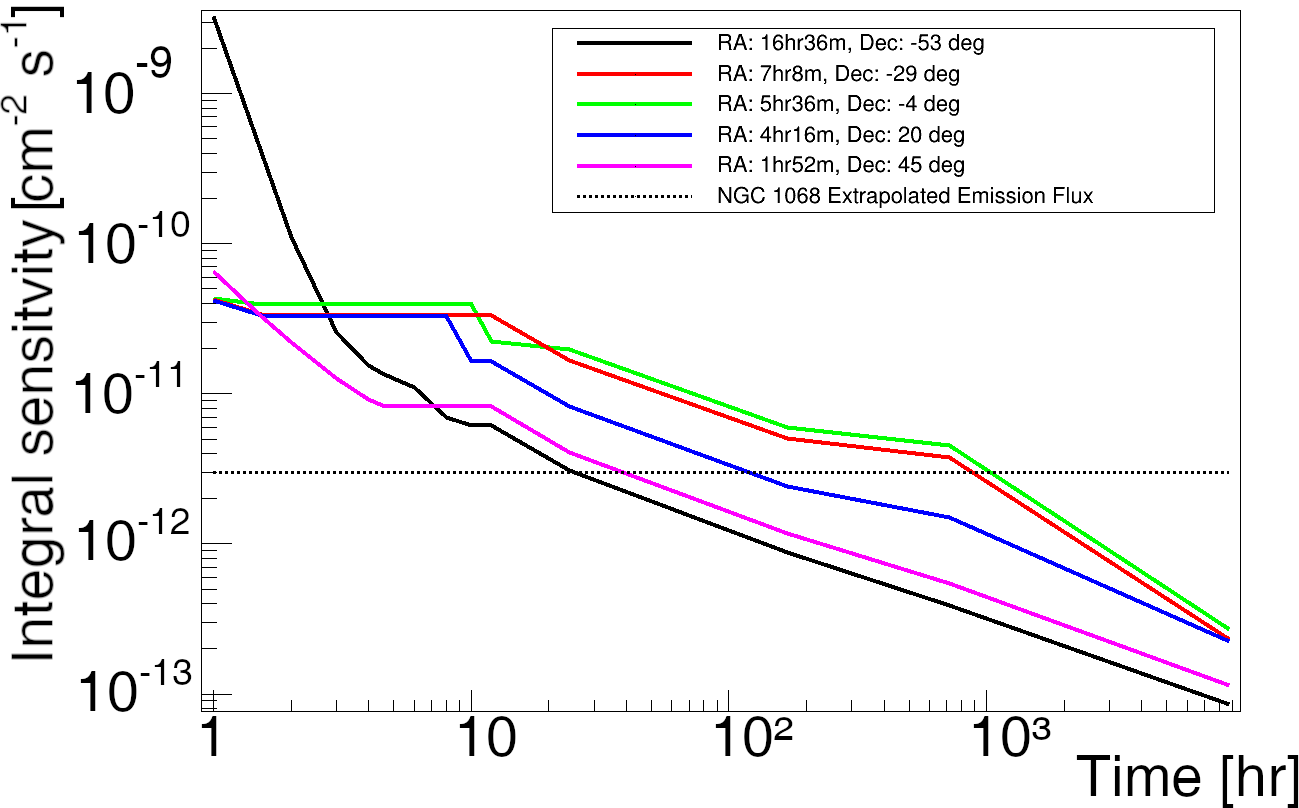}
\caption{Integrated point-source sensitivity versus observation time for the five source locations depicted in Figure~\ref{fig:transitsens}. The left panel assumes a source spectral index of $-2.2$. The data point shows the expected UHE flux extrapolated from the IceCube observation of neutrinos from the direction of TXS 0506+056 during the same 158 days \cite{ICTXS} interval as simulated. The right panel assumes a spectral index of $-3.2$. The dashed horizontal line shows the extrapolated neutrino flux from the direction of NGC 1068 \cite{Aartsen2020}.}
\label{fig:ptsens}
\end{figure}

Figure~\ref{fig:ptsens} shows the calculated sensitivity integrated from $10^6$\,GeV to $10^{10}$\,GeV for all flare durations and all five source locations. The sensitivity is best for the locations at the highest and lowest accessible declinations and about ten times worse at intermediate declinations. Choosing an observatory site at a different latitude will result in a shift of the highest-sensitivity regions. Differences in how the sensitivity changes with flare duration for the different source locations can also be attributed to differences in source observability. For a site at Frisco Peak, UT, simulated here, the TA hotspot, Cygnus A, and the historic blazars MRK 421 and MRK 501 are at declinations, which would be observable with the best sensitivity (see Figure~\ref{fig:accskymaps})

In order to put the simulated point-source sensitivities into contemporary context, we extrapolate the high-energy neutrino flux from TXS 0506+056 claimed by IceCube over a period of 158 days into the UHE range. We used the IceCube published power-law spectrum with flux normalization  $\Phi = 1.6_{-0.6}^{+0.7} \times 10^{-15} \text{ TeV}^{-1} \text{cm}^{-2} \text{s}^{-1}$ at 100 TeV and a spectral index of $\gamma = 2.2 \pm 0.2$ \cite{ICTXS} and integrate the flux from $10^6$\,GeV and $10^{10}$\,GeV to arrive at the data point in Figure \ref{fig:transitsens}. Although a naive extrapolation, it still demonstrates \emph{Trinity}'s sensitivity to UHE-neutrino sources of comparable power like those currently accessible in the HE-neutrino band. The full \emph{Trinity} Observatory, \emph{i.e.}, three sites instead of one, would have the sensitivity to detect a similar UHE neutrino emission as long as it is in Trinity's field of view.

In a second case study, we compare \emph{Trinity}'s performance against the IceCube-observed excess from the direction of NGC 1068. Using ten years of data from 2006 to 2018, IceCube found a best-fit flux of $\Phi \sim 3 \times 10^{-8} (E_\nu/\text{TeV})^{-3.2} \text{ GeV}^{-1} \text{cm}^{-2} \text{s}^{-1}$, where $E_{\nu}$ is neutrino energy \cite{Aartsen2020}. Extrapolating that spectrum to the $10^6$ GeV - $10^{10}$ GeV range yields an integral flux of of $\sim 3 \times 10^{-12} \text{ cm}^{-2} \text{s}^{-1}$. \emph{Trinity} would detect the extrapolated flux irrespective of the source declination within one week.

\section{Discussion and Outlook}
The UHE-neutrino band promises to be an exciting new window to the Universe provided new instruments are sensitive enough. The \emph{Trinity} Observatory is a planned system of air-shower imaging telescopes distributed at three different sites that promises to deliver the necessary sensitivity.

We presented updated diffuse-flux sensitivity calculations for \emph{Trinity} with a new neutrino and tau interaction and propagation model. Switching to NuTauSim has revealed an underestimation of the low-energy sensitivity of \emph{Trinity} in our previous calculations. The results presented here show that one single \emph{Trinity} telescope will detect the IceCube diffuse neutrino flux in five years, provided the spectrum does not turn over. The complete observatory has the sensitivity to discriminate between a continuation and a sharp turnover of the spectrum at $10^6$\,GeV.

We also calculated the point-source sensitivity for one \emph{Trinity} site. Depending on declination, a source remains  in the field-of-view of one \emph{Trinity} site between 1 hour and 12 hours. Transient sources with a hard spectrum and a flux of a few $10^{-12}$\,cm$^{-2}$s$^{-1}$ can be detected within one hour, whereas a flux of $10^{-14}$\,cm$^{-2}$s$^{-1}$ is detectable within one year.

The results demonstrate \emph{Trinity}'s high sensitivity as a UHE-neutrino detector, which will need to be validated with prototype detectors in the coming years.

\bibliographystyle{plain}
\bibliography{srcs}

%
%
%

\end{document}